\documentclass[12pt]{article}

\usepackage{amsmath,bbm,latexsym}

\newcommand{\be}{\begin{equation}}
\newcommand{\ee}{\end{equation}}
\newcommand{\ba}{\begin{eqnarray}}
\newcommand{\ea}{\end{eqnarray}}
\newcommand{\no}{\nonumber\\}

\begin{document}

\title{
\normalsize \hfill CFTP/10-001 \\[8mm]
\LARGE Renormalization-group constraints on Yukawa \\
alignment in multi-Higgs-doublet models
}

\author{P.M.~Ferreira,$^{(1,2)}$\thanks{E-mail: ferreira@cii.fc.ul.pt} \
L.~Lavoura,$^{(3)}$\thanks{E-mail: balio@cftp.ist.utl.pt} \
and Jo\~{a}o P.~Silva$\, ^{(1,3)}$\thanks{E-mail: jpsilva@cftp.ist.utl.pt}
\\*[3mm]
\small $^{(1)}$ Instituto Superior de Engenharia de Lisboa \\
\small 1959-007 Lisboa, Portugal
\\*[2mm]
\small $^{(2)}$ Centro de F\'\i sica Te\'orica e Computacional,
Universidade de Lisboa \\
\small 1649-003 Lisboa, Portugal
\\*[2mm]
\small $^{(3)}$ Centro de F\'\i sica Te\'orica de Part\'\i culas,
Instituto Superior T\'ecnico \\
\small 1049-001 Lisboa, Portugal
}

\date{14 January 2010}

\maketitle

\begin{abstract}
We write down the renormalization-group equations
for the Yukawa-coupling matrices in a general multi-Higgs-doublet model.
We then assume that the matrices of the Yukawa couplings
of the various Higgs doublets
to right-handed fermions of fixed quantum numbers
are all proportional to each other.
We demonstrate that,
in the case of the two-Higgs-doublet model,
this proportionality
is preserved by the renormalization-group running only in the cases
of the standard type-I,
II,
X,
and Y
models.
We furthermore show that a similar result holds
even when there are more than two Higgs doublets:
the Yukawa-coupling matrices to fermions of a given electric charge
remain proportional under the renormalization-group running
if and only if there is a basis for the Higgs doublets in which
all the fermions of a given electric charge
couple to only one Higgs doublet.
\end{abstract}

\newpage

The standard model of the electroweak interactions
has an extension with two Higgs doublets,
called the two-Higgs-doublet model (THDM).
In that extension the Yukawa couplings are given by
\be
\label{yuk}
\mathcal{L}_\mathrm{Y} =
- \bar Q_L \left[
\left( \Gamma_1 \phi_1 + \Gamma_2 \phi_2 \right) n_R
+
\left( \Delta_1 \tilde \phi_1 + \Delta_2 \tilde \phi_2 \right) p_R
\right]
- \bar L_L \left( \Pi_1 \phi_1 + \Pi_2 \phi_2 \right) \ell_R
+ \mathrm{H.c.},
\ee
where the $\phi_k$ ($k = 1, 2$) are the Higgs doublets,
$\tilde \phi_k \equiv i \tau_2 \phi_k^\ast$,
and $Q_L$,
$L_L$,
$n_R$,
$p_R$,
and $\ell_R$ are 3-vectors
in flavour space; $\Gamma_k$,
$\Delta_k$,
and $\Pi_k$
are the $3 \times 3$ complex matrices of the Yukawa couplings
to the right-handed down-type quarks,
up-type quarks,
and charged leptons,
respectively.

Recently~\cite{pich},
it has been proposed that the Yukawa-coupling matrices
may be proportional to each other:
\begin{subequations}
\label{align}
\ba
\Gamma_2 &=& d \Gamma_1, \label{aligngamma} \\
\Delta_2 &=& u \Delta_1, \label{aligndelta} \\
\Pi_2 &=& e \Pi_1, \label{alignpi}
\ea
\end{subequations}
where $d$,
$u$,
and $e$ are complex numbers.
This `Yukawa alignment'
might be the result of some symmetry and,
as such,
be stable under renormalization.
The purpose of this paper
is to investigate that possibility
by making use of the renormalization-group equations (RGE).

The one-loop RGE
for a model with $n$ Higgs doublets
have been presented in~\cite{grimus}
for the special case in which $\mathcal{L}_\mathrm{Y}$
only includes the matrices $\Pi_k$.
It is easy to extend them to the most general case.
Let $\mu$ be the renormalization scale and
$\mathcal{D} \equiv 16 \pi^2 \mu
\left( \mathrm{d} / \mathrm{d} \mu \right)$.
Then,
\begin{subequations}
\label{D}
\ba
\mathcal{D} \Gamma_k &=& a_\Gamma \Gamma_k
\label{aGamma} \\ & &
+ \sum_{l=1}^n \left[
3\, \mathrm{tr} \left( \Gamma_k \Gamma_l^\dagger
+ \Delta_k^\dagger \Delta_l \right)
+ \mathrm{tr} \left(\Pi_k \Pi_l^\dagger \right) \right] \Gamma_l
\label{bGamma} \\ & &
+ \sum_{l=1}^n \left(
- 2\, \Delta_l \Delta_k^\dagger \Gamma_l
+ \Gamma_k \Gamma_l^\dagger \Gamma_l
+ \frac{1}{2}\, \Delta_l \Delta_l^\dagger \Gamma_k
+ \frac{1}{2}\, \Gamma_l \Gamma_l^\dagger \Gamma_k \right),
\label{cGamma}
\ea
\end{subequations}
\begin{subequations}
\label{DDelta}
\ba
\mathcal{D} \Delta_k &=& a_\Delta \Delta_k
\label{aDelta} \\ & &
+ \sum_{l=1}^n \left[
3\, \mathrm{tr} \left( \Delta_k \Delta_l^\dagger
+ \Gamma_k^\dagger \Gamma_l \right)
+ \mathrm{tr} \left(\Pi_k^\dagger \Pi_l \right) \right] \Delta_l
\label{bDelta} \\ & &
+ \sum_{l=1}^n \left(
- 2\, \Gamma_l \Gamma_k^\dagger \Delta_l
+ \Delta_k \Delta_l^\dagger \Delta_l
+ \frac{1}{2}\, \Gamma_l \Gamma_l^\dagger \Delta_k
+ \frac{1}{2}\, \Delta_l \Delta_l^\dagger \Delta_k \right),
\label{cDelta}
\ea
\end{subequations}
\begin{subequations}
\label{DPi}
\ba
\mathcal{D} \Pi_k &=& a_\Pi \Pi_k
\label{aPi} \\ & &
+ \sum_{l=1}^n \left[
3\, \mathrm{tr} \left( \Gamma_k \Gamma_l^\dagger
+ \Delta_k^\dagger \Delta_l \right)
+ \mathrm{tr} \left(\Pi_k \Pi_l^\dagger \right) \right] \Pi_l
\label{bPi} \\ & &
+ \sum_{l=1}^n \left(
\Pi_k \Pi_l^\dagger \Pi_l
+ \frac{1}{2}\, \Pi_l \Pi_l^\dagger \Pi_k \right),
\label{cPi}
\ea
\end{subequations}
where $\mathrm{tr}$ denotes the trace and
the factors $3$ in~(\ref{bGamma}),
(\ref{bDelta}),
and (\ref{bPi}) are from colour.
Furthermore,
\begin{subequations}
\label{a}
\ba
a_\Gamma &=& - 8 g_s^2 - \frac{9}{4}\, g^2 - \frac{5}{12}\, {g^\prime}^2, \\
a_\Delta &=& - 8 g_s^2 - \frac{9}{4}\, g^2 - \frac{17}{12}\, {g^\prime}^2, \\
a_\Pi &=& - \frac{9}{4}\, g^2 - \frac{15}{4}\, {g^\prime}^2,
\ea
\end{subequations}
where $g_s$,
$g$,
and $g^\prime$ are the gauge coupling constants of $SU(3)_\mathrm{colour}$,
$SU(2)$,
and $U(1)$,
respectively.\footnote{The normalization of $g^\prime$ is such that
the $\phi_k$ have hypercharge $1/2$.}
We emphasize that equations~(\ref{D})--(\ref{a}) are valid for a model
with an \emph{arbitrary} number of Higgs doublets
and for any number of fermion generations.

Let us examine the contribution~(\ref{bGamma}) in the THDM
under the alignment hypothesis~(\ref{align}).
One obtains
\begin{subequations}
\label{DGamma}
\ba
\mathcal{D} \Gamma_1 &=& \left[
\left( 3 + 3 \left| d \right|^2 \right)
\mathrm{tr} \left( \Gamma_1 \Gamma_1^\dagger \right)
+ \left( 3 + 3 u d \right)
\mathrm{tr} \left( \Delta_1^\dagger \Delta_1 \right)
\right. \no & & \left.
+ \left( 1 + e^\ast d \right)
\mathrm{tr} \left( \Pi_1 \Pi_1^\dagger \right)
\right] \Gamma_1 + \cdots,
\\
\mathcal{D} \Gamma_2 &=& \left[
\left( 3 \left| d \right|^2 d + 3 d \right)
\mathrm{tr} \left( \Gamma_1 \Gamma_1^\dagger \right)
+ \left( 3 \left| u \right|^2 d + 3 u^\ast \right)
\mathrm{tr} \left( \Delta_1^\dagger \Delta_1 \right)
\right. \no & & \left.
+ \left( \left| e \right|^2 d + e \right)
\mathrm{tr} \left( \Pi_1 \Pi_1^\dagger \right)
\right] \Gamma_1 + \cdots.
\ea
\end{subequations}
We now require that the hypothesis~(\ref{align})
be stable under the RGE,
\textit{i.e.}\ we require
\begin{subequations}
\label{prop}
\ba
\mathcal{D} \Gamma_2 &=& d \left( \mathcal{D} \Gamma_1 \right),
\label{propGamma} \\
\mathcal{D} \Delta_2 &=& u \left( \mathcal{D} \Delta_1 \right),
\label{propDelta} \\
\mathcal{D} \Pi_2 &=& e \left( \mathcal{D} \Pi_1 \right).
\label{propPi}
\ea
\end{subequations}
Using~(\ref{DGamma}),
one finds that~(\ref{propGamma})
leads to
\begin{subequations}
\label{cond}
\ba
\left( u^\ast - d \right) \left( 1 + u d \right) &=& 0,
\label{conda} \\
\left( e - d \right) \left( 1 + e^\ast d \right) &=& 0.
\label{condb}
\ea
\end{subequations}
It is easy to check that the condition~(\ref{conda})
ensures that~(\ref{propGamma})
also holds for the terms in~(\ref{cGamma}).
One may furthremore check
that~(\ref{propDelta}) and~(\ref{propPi})
do not yield any further conditions
on $d$,
$u$,
and $e$ beyond~(\ref{cond}).
We have thus found that the alignment hypothesis
is (one-loop) renormalization-group invariant only
in one of the four cases
\begin{subequations}
\label{cases}
\ba
& & d = e = u^\ast, \label{1} \\
& & d = e = - \frac{1}{u}, \label{2} \\
& & d^\ast = u = - \frac{1}{e}, \label{3} \\
& & u = e^\ast = - \frac{1}{d}. \label{4}
\ea
\end{subequations}

Instead of using the basis $\left( \phi_1, \phi_2 \right)^T$
for the two Higgs doublets,
one may use any other basis,
related to that one through an $SU(2)$ transformation.
In particular,
one may use any of the following three bases:
\begin{subequations}
\label{bases}
\ba
\left( \begin{array}{c} \phi_{d1} \\ \phi_{d2} \end{array} \right)
&=& \left( 1 + \left| d \right|^2 \right)^{-1/2}
\left( \begin{array}{cc} 1 & d \\ - d^\ast & 1 \end{array} \right)
\left( \begin{array}{c} \phi_1 \\ \phi_2 \end{array} \right),
\label{basisd} \\
\left( \begin{array}{c} \phi_{u1} \\ \phi_{u2} \end{array} \right)
&=& \left( 1 + \left| u \right|^2 \right)^{-1/2}
\left( \begin{array}{cc} 1 & u^\ast \\ - u & 1 \end{array} \right)
\left( \begin{array}{c} \phi_1 \\ \phi_2 \end{array} \right),
\label{basisu} \\
\left( \begin{array}{c} \phi_{e1} \\ \phi_{e2} \end{array} \right)
&=& \left( 1 + \left| e \right|^2 \right)^{-1/2}
\left( \begin{array}{cc} 1 & e \\ - e^\ast & 1 \end{array} \right)
\left( \begin{array}{c} \phi_1 \\ \phi_2 \end{array} \right).
\label{basise}
\ea
\end{subequations}
If
the alignment conditions~(\ref{align}) hold,
then the Yukawa Lagrangian~(\ref{yuk}) may be rewritten
\be
\label{yuk2}
\mathcal{L}_\mathrm{Y} =
- \bar Q_L \left(
\phi_{d1} \sqrt{1 + \left| d \right|^2} \Gamma_1 n_R
+
\tilde \phi_{u1} \sqrt{1 + \left| u \right|^2}\Delta_1 p_R
\right)
- \bar L_L \phi_{e1} \sqrt{1 + \left| e \right|^2} \Pi_1 \ell_R
+ \mathrm{H.c.},
\ee
where $\tilde \phi_{u1} \equiv i \tau_2 \phi_{u1}^\ast$.
The various cases~(\ref{cases})
may then be interpreted in the following way:
\begin{itemize}
\item In the case~(\ref{1}),
$\phi_{d1} = \phi_{u1} = \phi_{e1}$,
hence
\[
\frac{\mathcal{L}_\mathrm{Y}}{\sqrt{1 + \left| d \right|^2}} =
- \bar Q_L \left(
\phi_{d1} \Gamma_1 n_R
+
\tilde \phi_{d1} \Delta_1 p_R
\right)
- \bar L_L \phi_{d1} \Pi_1 \ell_R
+ \mathrm{H.c.},
\]
which means that,
in the basis $\left( \phi_{d1}, \phi_{d2} \right)$,
only $\phi_{d1}$ has Yukawa couplings.
This is called the type-I THDM~\cite{typeI}.
\item In the case~(\ref{2}),
\[
\frac{\mathcal{L}_\mathrm{Y}}{\sqrt{1 + \left| d \right|^2}} =
- \bar Q_L \left(
\phi_{d1} \Gamma_1 n_R
-
\tilde \phi_{d2}\, \frac{\Delta_1}{d}\, p_R
\right)
- \bar L_L \phi_{d1} \Pi_1 \ell_R
+ \mathrm{H.c.},
\]
which means that only $\phi_{d1}$ has Yukawa couplings
to the down-type quarks and to the charged leptons,
while only $\phi_{d2}$ has Yukawa couplings
to the up-type quarks.
This is called the type-II THDM~\cite{typeII}.
\item In the case~(\ref{3}),
\[
\frac{\mathcal{L}_\mathrm{Y}}{\sqrt{1 + \left| d \right|^2}} =
- \bar Q_L \left(
\phi_{d1} \Gamma_1 n_R
+
\tilde \phi_{d1} \Delta_1 p_R
\right)
+ \bar L_L \phi_{d2}\, \frac{\Pi_1}{d^\ast}\, \ell_R
+ \mathrm{H.c.},
\]
which means that only $\phi_{d1}$ has Yukawa couplings
to the quarks while only $\phi_{d2}$ has Yukawa couplings
to the charged leptons.
This is called the type-X THDM~\cite{type}.
\item In the case~(\ref{4})
\[
\frac{\mathcal{L}_\mathrm{Y}}{\sqrt{1 + \left| d \right|^2}} =
- \bar Q_L \left(
\phi_{d1} \Gamma_1 n_R
-
\tilde \phi_{d2}\, \frac{\Delta_1}{d}\, p_R
\right)
+ \bar L_L \phi_{d2}\, \frac{\Pi_1}{d^\ast}\, \ell_R
+ \mathrm{H.c.},
\]
which means that only $\phi_{d1}$ has Yukawa couplings
to the down-type quarks
while only $\phi_{d2}$ has Yukawa couplings
to the charged leptons
and to the up-type quarks.
This is called the type-Y THDM~\cite{type}.
\end{itemize}
It is well known that all these cases may be obtained through the
imposition of appropriate $\mathbbm{Z}_2$ symmetries
on the Yukawa Lagrangian~(\ref{yuk}).

We proceed to analyze the case of an arbitrary number $n$ of Higgs doublets,
where
\be
\label{general_yuk}
\mathcal{L}_\mathrm{Y} = - \bar Q_L
\sum_{k=1}^n \left( \Gamma_k \phi_k n_R +
\Delta_k \tilde \phi_k p_R \right)
- \bar L_L \sum_{k=1}^n \Pi_k \phi_k \ell_R
+ \mathrm{H.c.},
\ee
with a generalized alignment hypothesis
\begin{subequations}
\label {general_align}
\ba
\Gamma_k &=& d_k \Gamma_1, \\
\Delta_k &=& u_k \Delta_1, \\
\Pi_k &=& e_k \Pi_1,
\ea
\end{subequations}
where $d_1=u_1=e_1=1$.
Inserting the conditions~(\ref{general_align})
in equation~(\ref{D}) leads to
\ba
\mathcal{D} \Gamma_k
&=&
a_\Gamma\, d_k\, \Gamma_1
\nonumber\\
& &
+ \sum_{l=1}^n \left[ 3 d_k \left| d_l \right|^2
\mathrm{tr} \left( \Gamma_1 \Gamma_1^\dagger \right)
+ 3 u_k^\ast u_l d_l\,
\mathrm{tr} \left( \Delta_1^\dagger \Delta_1 \right)
+ e_k e_l^\ast d_l\, 
\mathrm{tr} \left( \Pi_1 \Pi_1^\dagger \right) \right]
\Gamma_1
\no & &
+ \sum_{l=1}^n \left[
\left( - 2 u_l u_k^\ast d_l + \frac{1}{2} \left| u_l \right|^2 d_k \right)
\Delta_1 \Delta_1^\dagger \Gamma_1
+ \frac{3}{2}\,
|d_l|^2 d_k 
\Gamma_1 \Gamma_1^\dagger \Gamma_1 \right].
\ea
It follows that
\ba
\mathcal{D} \Gamma_k - d_k \mathcal{D} \Gamma_1
&=&
\sum_{l=1}^n \left[
3 \left( u_k^\ast - d_k \right)
u_l d_l\, \mathrm{tr} \left( \Delta_1^\dagger \Delta_1 \right)
+ \left( e_k - d_k \right) e_l^\ast d_l\,
\mathrm{tr} \left( \Pi_1 \Pi_1^\dagger \right) \right] \Gamma_1
\nonumber\\
& &
- 2 \left( u_k^\ast - d_k \right)
\sum_{l=1}^n u_l d_l 
\Delta_1 \Delta_1^\dagger \Gamma_1.
\label{quant}
\ea
We require the quantity~(\ref{quant}) to vanish for all $k$.
A similar requirement is imposed on
$\mathcal{D} \Delta_k - u_k \mathcal{D} \Delta_1$
and on $\mathcal{D} \Pi_k - e_k \mathcal{D} \Pi_1$.
We find that the generalized alignment hypothesis~(\ref{general_align})
is preserved by the renormalization-group evolution if and only if
\begin{subequations}
\label{gen}
\ba
\left( u_k^\ast - d_k \right) \sum_{l=1}^n u_l d_l &=& 0,
\label{gen_ud} \\
\left( e_k - d_k \right) \sum_{l=1}^n e_l^\ast d_l &=& 0,
\label{gen_ed} \\
\left( u_k^\ast - e_k \right) \sum_{l=1}^n u_l e_l &=& 0
\label{gen_ue}
\ea
\end{subequations}
are satisfied for all $k$.
Now,
if the generalized alignment hypothesis holds,
then the Yukawa Lagrangian~(\ref{general_yuk}) may be rewritten as
\be
\label{general_yuk_2}
\mathcal{L}_\mathrm{Y} =
- \bar Q_L \left( N_d \phi_{d1} \Gamma_1 n_R
+ N_u \tilde \phi_{u1} \Delta_1 p_R \right)
- \bar L_L N_{e} \phi_{e1} \Pi_1 \ell_R
+ \mathrm{H.c.},
\ee
where
\begin{subequations}
\label{phi_diagonal}
\ba
N_d \phi_{d1} = \sum_{k=1}^n d_k \phi_k, & &
N_d = \sqrt{\sum_{k=1}^n \left| d_k \right|^2},
\\
N_u \phi_{u1} = \sum_{k=1}^n u_k^\ast \phi_k, & &
N_u = \sqrt{\sum_{k=1}^n \left| u_k \right|^2},
\\
N_e \phi_{e1} = \sum_{k=1}^n e_k \phi_k, & &
N_e = \sqrt{\sum_{k=1}^n \left| e_k \right|^2}.
\ea
\end{subequations}
Thus,
condition~(\ref{gen_ud}) means that either $\phi_{u1} = \phi_{d1}$
or else $\phi_{u1}$ and $\phi_{d1}$ are orthogonal;
the conditions~(\ref{gen_ed}) and (\ref{gen_ue})
have analogous interpretations.
This finally leaves us with only five possibilities:
\begin{enumerate}
\item \ $\phi_{d1} = \phi_{u1} = \phi_{e1}$; \label{um}
\item \ $\phi_{u1}$ is orthogonal to $\phi_{d1} = \phi_{e1}$; \label{dois}
\item \ $\phi_{e1}$ is orthogonal to $\phi_{d1} = \phi_{u1}$; \label{tres}
\item \ $\phi_{d1}$ is orthogonal to $\phi_{e1} = \phi_{u1}$; \label{quatro}
\item \ $\phi_{d1}$,
$\phi_{u1}$,
and $\phi_{e1}$ are all orthogonal to each other. \label{cinco}
\end{enumerate}
The possibilities~\ref{um}--\ref{quatro} reproduce
the four cases~(\ref{1})--(\ref{4}) already present in the THDM.
The possibility~\ref{cinco} is new:
if there are three or more Higgs doublets,
it is possible that each charged-fermion sector
couples to a separate
Higgs doublet---this is of course impossible in the THDM.

It is easy to show that all
five cases~\ref{um}--\ref{cinco}
may be enforced through simple $\mathbbm{Z}_2$ symmetries.

To summarize,
in this paper we have shown that
Yukawa alignment in the two-Higgs-doublet model,
as proposed in~\cite{pich},
can only be stable under the renormalization group
in the well-known cases
in which only one Higgs doublet has Yukawa couplings
to the right-handed fermions of each electric charge.
All other cases of Yukawa alignment
are not consistent with the renormalization group,
which means that,
either they cannot be enforced by any symmetries,
or they can be enforced by symmetries only in the context
of a model larger than the THDM.
We have generalized the notion of alignment
to models with an arbitrary number of Higgs doublets
and have shown that,
also in that generalization,
alignment is only stable under the RGE when
the right-handed fermions of each electric charge
all couple to only one Higgs doublet.

\paragraph{Acknowledgements:}
We thank A.\ Pich and P.\ Tuz\'{o}n for reading and commenting
on the manuscript.
The work of P.M.F. is supported in part by the Portuguese
\textit{Funda\c{c}\~{a}o para a Ci\^{e}ncia e a Tecnologia} (FCT)
under contract PTDC/FIS/70156/2006.
The work of L.L.\ and of J.P.S.\ is funded by FCT
through the project U777--Plurianual.

\end{document}